\documentclass[a4paper,conference]{IEEEtran}

\ifCLASSINFOpdf
\else
\fi
\usepackage{url}

\usepackage[english]{babel}

\usepackage{xcolor}

\usepackage{soul}

\usepackage[textsize=tiny,colorinlistoftodos]{todonotes}

\usepackage[
        bibstyle=ieee,
        citestyle=numeric-comp,
        maxcitenames=6,
        doi=false,
        eprint=false
    ]{biblatex}
        
\AtEveryBibitem{
  \clearfield{urlyear}%
  \clearlist{language}%
 }

\makeatletter
    \define@key{todonotes}{sm}[]{
        \setkeys{todonotes}{author=\textbf{Sergiy}, color=green!30}}%
\makeatother

\usepackage[range-phrase={\,--\,}, range-units=single, per-mode=symbol, detect-all]{siunitx}

\makeatletter
  \@addtoreset{paragraph}{section}
\makeatother

\usepackage[acronym]{glossaries}
\newacronym{vr}{VR}{Virtual Reality}
\newacronym{xr}{XR}{eXtended Reality}
\newacronym{qoe}{QoE}{Quality of Experience}
\newacronym{qos}{QoS}{quality of service}
\newacronym{uav}{UAV}{unmanned aerial vehicle}
\newacronym{faas}{FaaS}{Function-as-a-Service}
\newacronym{pubsub}{Pub/Sub}{Publish/Subscribe}
\newacronym{h3d}{H3D}{holographic 3D}
\newacronym{ai}{AI}{artificial intelligence}
\newacronym{nbmp}{NBMP}{network-based media processing}
\newacronym{kpi}{KPI}{key performance indicator}
\newacronym{hc}{\emph{HoloCom}}{\emph{Holographic Communication}}
\newacronym{ml}{ML}{machine learning}
\newacronym{rem}{REM}{radio environment map}
\newacronym{drl}{DRL}{deep reinforcement learning}
\newacronym{sdr}{SRD}{software-defined radio}
\newacronym[firstplural=radio access technologies]{rat}{RAT}{radio access technology}

\usepackage{balance}

\usepackage{cleveref}

\usepackage[all]{nowidow}

\begin{document}
%

\title{6G NeXt --- Towards 6G Split Computing Network Applications: Use Cases and Architecture}



%
\author{
    \IEEEauthorblockN{
        Sergiy Melnyk\IEEEauthorrefmark{1}, Qiuheng Zhou\IEEEauthorrefmark{1}, Hans D. Schotten\IEEEauthorrefmark{1}\IEEEauthorrefmark{2},
        Mandy Galkow-Schneider\IEEEauthorrefmark{3}, Ingo Friese\IEEEauthorrefmark{3}, \\
        Tobias Pfandzelter\IEEEauthorrefmark{4}, David Bermbach\IEEEauthorrefmark{4},
        Louay Bassbouss\IEEEauthorrefmark{5},
        Alexander Zoubarev\IEEEauthorrefmark{5},
        Andy Neparidze\IEEEauthorrefmark{5}, \\
        Arndt Kritzner\IEEEauthorrefmark{6}, 
        Enrico Zschau\IEEEauthorrefmark{7},        
        Prasenjit Dhara\IEEEauthorrefmark{8},
        Steve Göring\IEEEauthorrefmark{8},
        William Menz\IEEEauthorrefmark{8}
        Alexander Raake\IEEEauthorrefmark{8},\\
        Wolfgang Rüther-Kindel\IEEEauthorrefmark{9},
        Fabian Quaeck\IEEEauthorrefmark{9},
        Nick Stuckert\IEEEauthorrefmark{9}, 
        Robert Vilter\IEEEauthorrefmark{9},
    }
    \IEEEauthorblockA{
        \IEEEauthorrefmark{1}
        Intelligent Networks,
        German Research Center for Artificial Intelligence (DFKI GmbH), Kaiserslautern, Germany \\
        Email: \{sergiy.melnyk, qiuheng.zhou, hans\_dieter.schotten\}@dfki.de
    }
    \IEEEauthorblockA{
        \IEEEauthorrefmark{2}
        Institute for Wireless Communication and Navigation,
        University of Kaiserslautern-Landau, Kaiserslautern, Germany\\
        Email: schotten@rptu.de
    }
    \IEEEauthorblockA{
        \IEEEauthorrefmark{3}
        Deutsche Telekom AG, Berlin, Germany \\
        Email: \{mandy.galkow-schneider, ingo.friese\}@telekom.de
    }
    \IEEEauthorblockA{
        \IEEEauthorrefmark{4}
        Mobile Cloud Computing Group, Technische Universität Berlin \& ECDF, Berlin, Germany \\
        Email: \{tp, db\}@mcc.tu-berlin.de
    }
    \IEEEauthorblockA{
        \IEEEauthorrefmark{5}
        Fraunhofer FOKUS, Berlin, Germany \\
        Email: \{louay.bassbouss, alexander.zoubarev, andy.neparidze\}@fokus.fraunhofer.de
    }
    \IEEEauthorblockA{
        \IEEEauthorrefmark{6}
        Logic Way GmbH, Schwerin, Germany \\
        Email: kritzner@logicway.de
    }
    \IEEEauthorblockA{
        \IEEEauthorrefmark{7}
        SeeReal Technologies GmbH, Dresden, Germany \\
        Email: ez@seereal.com
    }
    \IEEEauthorblockA{
        \IEEEauthorrefmark{8}
        Audiovisual Technology Group, Technische Universität Ilmenau, Ilmenau, Germany \\
        Email: \{prasenjit.dhara, steve.goering, 
        william.menz, alexander.raake\}@tu-ilmenau.de
    }
    \IEEEauthorblockA{
        \IEEEauthorrefmark{9}
        Aviation Engineering Group, Technical University of Applied Sciences Wildau, Wildau, Germany \\
        Email: \{wkindel, quaeck, stuckert, vilter\}@th-wildau.de
    }
}

\maketitle


\begin{abstract}
    The definition of the sixth generation mobile communication is in the full swing. Mobile communication aims for the convergence of physical, human and digital world.
    Research project \emph{6G NeXt} is considering two demanding use cases, holographic communications and drones anti-collision system, which set heterogeneous requirements on the communication as well as the computing infrastructure. In both use cases, the clients are widely spread in the network and are cooperatively interacting with each other. Especially for holographic communication, also high processing power is required. This makes a high-speed distributed backbone computing infrastructure, which realises the concept of split-computing, inevitable. Furthermore, tight integration between processing facilities and wireless network is required in order to provide adequate quality of service to the users. This paper illustrates the use case scenarios and their requirements. Afterwards, an appropriate solution approach to realise those is elaborated. Here, the novel technological approaches are discussed based on the developed overall communication and computing architecture.
    
\end{abstract}

\begin{refsection}
    \let\thefootnote\relax\footnote{This is a preprint of a work accepted but not yet published at the 27. ITG-Fachtagung Mobilkommunikation. Please cite as: \fullcite{melnyk20236g2}.}
\end{refsection}%


\section{Introduction}

The upcoming sixth generation of mobile networks is currently in the early development phase of discovering future-oriented services and applications, their requirements, as well as novel technologies to realize them. It is expected that the emerging amount of data, especially in the field of \gls{xr}, will be tremendous. Thus, the required data rates, transmission latency, and computation power will exceed far beyond the capabilities of existing 5G networks~\cite{eu5Gvision,itunetwork2030}. 

Furthermore, for the realization of such complex use cases, a sophisticated computing infrastructure is required. That is the core research topic of the project \emph{6G NeXt -- 6G Native Extensions for \gls{xr} Technologies}. The goal of the project is to extend the edge computing concept, distributing the processing facilities on the different levels of the backbone network~\cite{bermbach2018research}. Provided the tight integration with the communication infrastructure, the split computing approach would significantly enhance the 6G network experience~\cite{bassbouss20215g}. Based on two example use cases with heterogeneous and demanding requirements, the project \emph{6G NeXt} aims to develop and evaluate a joint connectivity and compute infrastructure, which will introduce new processing speeds in conjunction with highly dynamic geo-distributed computing capabilities within a mobile network.

One of the use cases will realise a fully holographic communication platform bringing today's video calls to the next level. This platform will provide a real 3D experience with natural eye contact to the users. This means that no bulky \gls{vr} glasses are required. The special holographic display will provide a natural viewing experience with properly placed holograms, which are sharp at any view depth. The second use case deals with air traffic management for \glspl{uav} as well as general aviation in a shared airspace, e.\,g., in vicinity of an airport. The goal is to provide a real-time capable anti-collision system for mixed traffic types. A dedicated simulation engine monitors the position of all assigned aircraft, predicts their flight paths, and detects the potential risk of collisions.


To realize the concept of split computing in 6G networks, a tight integration of the communication infrastructure with the computational facilities is required. Therefore, the whole wireless protocol stack from the application layer down to the physical layer should be optimized to provide flexibility and agility for dynamic resource re-allocation. \emph{6G NeXt} is considering the algorithms for agile communication link adaptation based on the application requirements on the one hand, and the distribution of the resource in the network on the other hand. Combined with real-time radio protocol profiles that can be automatically adapted to the requirements of the example applications, the efficiency, quality, and real-time capability of the services will be increased.

\section{Use Cases}

\subsection{\acrshort{uav} Collision Detection and Avoidance}

The idea of an anti-collision system for drones or \gls{uav} is derived from the research project VIGA (Virtual Instructor for General Aviation)~\cite{VIGA} of the Technical University of Applied Sciences Wildau. The chosen approach for the system's design is based upon the idea of a simulation of an aircraft that runs faster than the real-time. In this way, evaluations of the possible flight trajectory can be estimated and warnings about hazardous situations in the near future can be presented to the pilot on primary flight display for example on an artificial horizon in the cockpit.

In the \emph{6G NeXt} project, this approach is extended to \glspl{uav} in the use case \emph{Smart Drones}. Future outlooks for aviation forecast an increasing number of remotely operated flight missions as well as an increasing number of manned flights~\cite{VIGA}. Until today many strategies have been developed on the topic of how to implement \glspl{uav} in the civil aviation airspace. In consequence and in order to ensure a safe operation in a shared airspace, a common anti-collision system will be necessary in the future. 
The use case \emph{Smart Drones} focuses primarily on drone to drone anti-collision but future extensions to manned aviation are also conceivable. Therefore,
the goal is to provide a safe and reliable connectivity to any \gls{uav} operating in dangerous areas near airports and to provide the software layer for this dynamic application.
The novelty of the system lies in its different approach in comparison to today's anti-collision system.

To simulate the flight paths of each \gls{uav}, a reliable connectivity to a central ground station is necessary since the simulation is not carried out onboard the \gls{uav} but on a ground-based computing station. This is done due to the fact that the onboard computational resources are not sufficient and more resources lead to higher take off weight which in return is decreasing mission time or mission range. The ground station is meant to collect all data streams of each individual \gls{uav} flying in the specific sector. The transmitted data contains altitude, position, course and velocity information (state values) as well as intended mission profile data and \gls{uav}-specific data (flight model values). The digital representations (digital twins) of the \glspl{uav} are fed by this \gls{uav}-specific information and the simulation is carried out with respect to the currently received state values. It is planned that each flight path is then evaluated concurrently with paths of other \glspl{uav}. If a collision scenario is predicted by the system, evading trajectories will be calculated and simulated with the digital twins before the manoeuvre is transmitted and performed by the real \gls{uav}.

The anti-collision system will be based upon a simulation approach. The input parameters of the simulation are primarily driven by the \gls{uav}-specific data which is not expected to change during a flight and by the current inertial, satellite-based and air data which is acquired on the drone at a high frequency. This is why a low latency in the communication link is important, since the overall time of data transmission, simulation and manoeuvre execution has to be minimized. The more current the input data to the simulation is, the better the systems result will match with the real situation and improves the system's performance. 
    
Moreover, a wide range of low latency and highly reliable connectivity is essential for this mobile airborne application. 
The movement of the \glspl{uav} through different sectors require a new dynamic data distribution. This requires a software architecture that is capable of allocating the \gls{uav}'s data dynamically and hand over existing simulation data to other sectors.
    

\subsection{Holographic Communication (HoloCom)}

The idea and goal of the use case \gls{hc} is to research and develop a 3D video communication system based on true \gls{h3d} displays. We aim for a realistic and natural like remote communication between individuals enabled by \gls{h3d} technology invented by one of our partners. This technology creates light information in space the same way as real objects would do, to allow natural viewing for the human eye including continuous depth / eye-focus and convergence. This enables a natural like, audiovisual communication by reconstructing the call counterpart as live and true 3D hologram in space. No 3D-glasses are required due to integrated Eye-Tracking. Instead of using a \gls{h3d} display, VR headsets could be optionally used, of course with limited options to capture the persons head and face.

The \gls{hc} use case requires 3D capturing of real persons in real-time, performed on the client. The 3D data will be transformed into a 3D model to be synthesized into a virtual environment within the Edge domain. Rendered 3D views are finally transferred to the other client to be encoded into holograms and presented on a \gls{h3d} display. The rendered 3D views depend on the eye-location of the observer looking at the \gls{h3d} display, which results in en extremely low delay requirement to be tackled by 6G. But beside delay, also bandwidth requirements are high to satisfy demand for high quality person representation in full 3D. Some key research will also be done in the field of 3D capturing with several cameras of different types, from normal RGB sensors to time of flight or depth from stereo based depth capturing methods. This will be based on a technology of one of our partners, which already enables high quality 360 degree live scanning, producing more than 10 terabytes of date per minute.




The user experience is a central factor for real-time immersive media communication over the network. Traditional \gls{qos} metrics such as, packet loss, network delay, and round-trip-times, etc., are important for media communication, but it is not sufficient for user’s perceived satisfaction. Higher values of \gls{qos} do not necessarily ensure higher \gls{qoe} \cite{agiwal2016next}. User aspects must be included in the analysis of holographic 3D video communication as human perception is more sensitive to certain display errors. In the \gls{hc} use case,  the focus is on subjective as well as objective methods for the evaluation of the immersive, audiovisual communication systems. 
    
For \gls{qoe} in \gls{hc}, ground-truth data for model development is collected first. Here, the subjective measures include several questionnaires related to co-presence, social presence, trust, immersion, communication satisfaction, as well as perceived quality in terms of classical overall mean opinion score (MOS) assessment. In addition, conversational and behavioral data of the participants recorded during the experiment will be analyzed to indirectly evaluate user experience \cite{smith2018communication, skowronek2022quality}. Along with these user-based analyses, so-called objective measures will be computed to evaluate the 3D rendering, encoding and decoding of data representing the conversation partners.
        


\section{6G Communication and Computing Infrastructure}

The performance and efficiency of a new network generation will be determined by high-performance radio interfaces with application-optimized radio protocols as well as by ultra-fast software stacks, intelligent media processing and the deep integration of \gls{ai}  to optimize the overall system. The focus is on an implementation with open interfaces, easy integrability, sustainable development and optimized economic efficiency to achieve a broad social acceptance of the new 6G technology.  

In the following, the key technologies for communication and computing infrastructure are presented. Especially, a joint architecture, which combines communication and computing worlds, is proposed.

\subsection{Split Computing}

\begin{figure*}
    \centering
    \includegraphics[width=.8\linewidth]{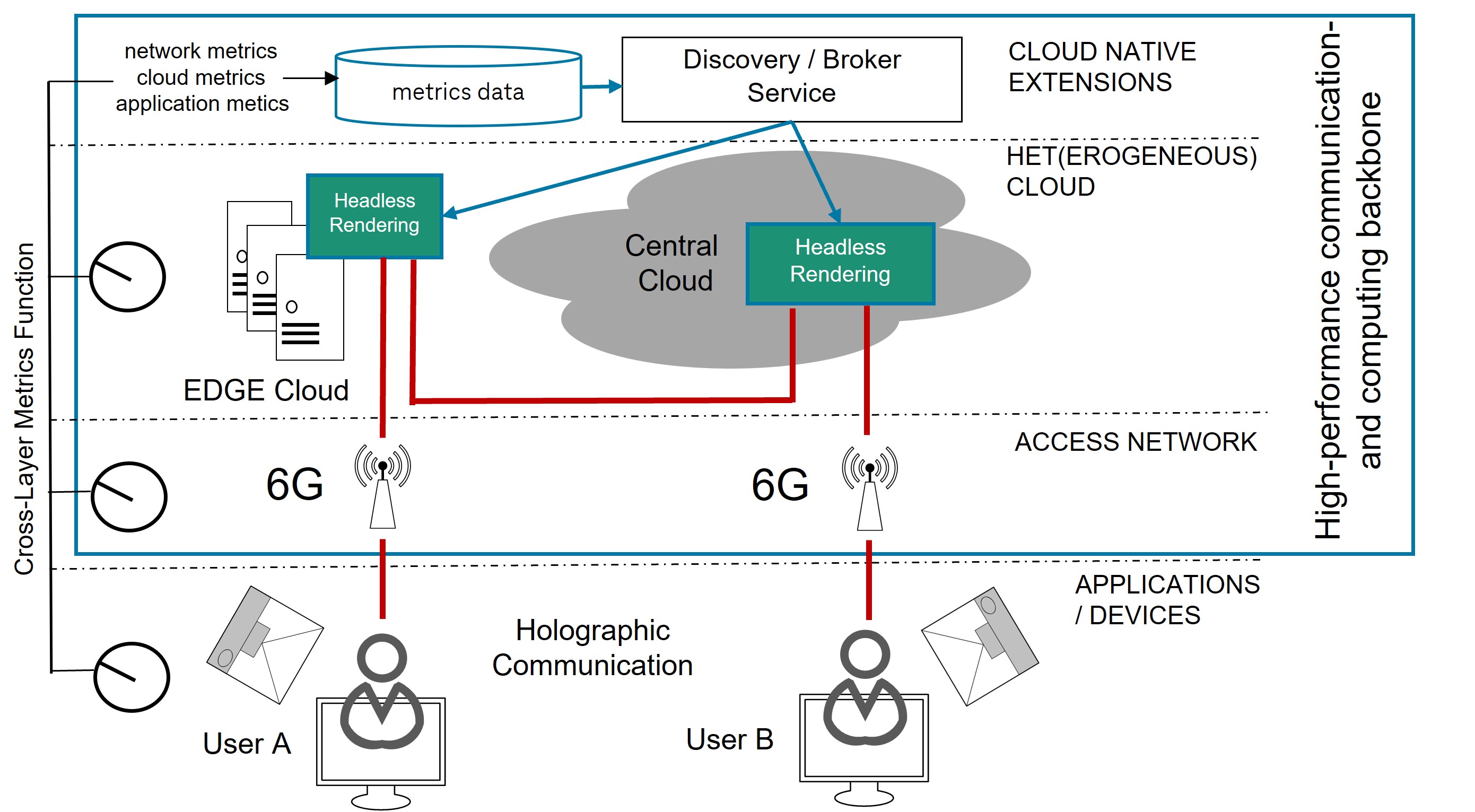}
    \caption{High-performance communication-and-computing backbone for holographic communication}
    \label{fig:architecture}
\end{figure*}

Distributing the computation power in the backbone network based on the required \glspl{kpi} enhances the concept of edge computing and provides new possibilities for the future applications. 

\subsubsection{Cross-Layer Metrics Function} \label{cross_layer_metrics}
A new metrics function is going to be implemented to constantly gain status and characteristics of network, computing layer and applications as shown in Figure~\ref{fig:architecture}. The data are used to enable applications to find the optimal environment to run in, to connect to, or to be deployed at. Thus computing capacities and connectivity can be used more efficiently. The collected data is also a source for \gls{ai} algorithms aiming at the optimization of the whole system.

\subsubsection{High-performance communication-and-computing backbone} 

One of the main objectives of the project is the development of a high-performance backbone layer, which allocates computing capacities and connectivity according to the requirements for latency, available CPU/GPU capacity, energy consumption and costs, among others. Additional services and extensions optimize cloud infrastructures commonly used today. These cloud-native extensions, the cloud layer as well as the network form the High-performance communication-and-computing backbone (see Figure~\ref{fig:architecture}).

A combined Discovery and Resource Broker Function helps clients and cloud-based services to find the required resources according to their needs based on measurement results of the cross-layer metrics function. The discovery service helps clients to find their optimal service. A holographic client for example might discover the cloud rendering service (also known as headless rendering) with the smallest available latency at the lowest price. This does not have to be the physically nearest rendering service. Because latency might be caused by the network or by the computing layer when there is not enough CPU or GPU capacity available. The rendering service might use the Resource Broker Function in case it needs to start a new service instance in order to increase its capacity. The broker is able to find available CPU/GPU computing capacity in conjunction with sufficient connectivity.

The classic separation of network and computing infrastructure in layers is eliminated and replaced by a modular approach. The clear distinction between computing infrastructures also starts to blur. The High-performance communication-and-computing backbone enables  a flexible split of computing tasks according to the requirements of the application. And applications get all information and support to decide where to run, where to connect or to be  deployed to. This approach complements the edge cloud specification in \emph{3GPP 5G}~\cite{3gpp.23.558} in the direction of a convergence of computing and connectivity across the whole cloud continuum.

\subsubsection{Split rendering and \acrshort{nbmp}} 
For the concept of split rendering, the possibility to define and distribute media processing workflows across the available infrastructure is a necessity, which holds especially true for the architecture we propose. To accommodate this requirement, we utilize the \gls{nbmp} standard \cite{iso_NBMP}, which defines software components and workflow description schemes facilitating the deployment and reuse of media processing workflows in a network. The specification describes a central controller entity and a repository storing textual representations of the available processing functions of which workflows are composed. Besides that, it offers a standardized scheme for describing functions and workflows, and defines the interfaces between the involved components. 

In a split rendering scenario, workflow information can be supplemented with processing or location requirements to achieve a distributed workflow deployment by the central controller. The controller can draw information from the resource broker mentioned in section \ref{cross_layer_metrics}, allowing for an intelligent distribution of media processing functions.
Due to refraining from specifying the underlying processing network, \gls{nbmp} is a perfect fit, not only for our architecture but for 6G in general, and from the two discussed use cases, greatly benefits the holographic communication by allowing an on-demand distributed deployment of a predefined processing pipeline. 

In a simplified pipeline consisting of preprocessing, video and audio processing, rendering, encoding, and decoding, the preprocessing step can be configured to always be executed on-site as close to the capturing equipment as possible. For the next three steps, increased computing resource requirements and proximity stipulations will lead the controller to deploy the corresponding processing functions to the nearest edge, while the final decoding step will be deployed close to the communication partner. Once defined, this processing workflow can be parameterized, reused and easily extended or modified due the standardized description schema, ensuring compatibility with any workflows adhering to the \gls{nbmp} standard.

\subsubsection{Edge-to-Cloud Serverless Platform}

The high-speed backbone and embedded compute resources are combined with powerful software abstractions for application services.
We adopt concepts of the fast-growing serverless cloud computing paradigm for \emph{computation}, \emph{data management}, and \emph{communication} on the edge-cloud continuum.
The focus is on making these abstractions adaptive and scalable, integrating geo-distributed edge resources with cloud bursting.

\paragraph{Compute}

The \gls{faas} paradigm has become increasingly popular in cloud computing, as it allows developers to build highly-scalable, cost-efficient applications by composing small, event-driven, stateless functions~\cite{pfandzelter2019iot}.
We adopt this paradigm to realize large-scale edge-to-cloud computing, as it has numerous advantages for edge compute infrastructure:
The small size of serverless functions and their ephemeral nature allow fine-grained dynamic resource allocation, without the need for long lead times and blocking resources with idle services.
The event-driven nature of serverless applications allows dynamic routing of requests based on invocation location, infrastructure utilization, failures in the network, and other parameters.
A key challenge in bringing \gls{faas} to the edge is in decreasing the performance efficiency overhead of its high level of abstraction.
Edge \gls{faas} platforms, such as \emph{tinyFaaS}~\cite{paper_pfandzelter_tinyfaas}, are designed to be lightweight enough to run even on constrained edge devices.


\paragraph{Data}

Stateless functions alone are insufficient to build many of the applications we target in this project.
For example, \gls{uav} collision avoidance requires state in the form of trajectory and capability information for \glspl{uav}.
As a result, we also develop abstractions around data management that are tightly integrated with our other components.
The migration of functions in the compute layer is assisted by automated data migration across the geo-distributed edge-to-cloud infrastructure~\cite{hasenburg2019fbase,hasenburg2020towards}.
Application-controlled data replica placement, leveraging building blocks such as \emph{FReD}~\cite{pfandzelter2023fred}, is combined with platform-side autonomous strategies for data placement and migration to reduce access latency in order to reach \gls{qos}/\gls{qoe} targets.
Geo-distributed serverless function code can then access local data copies transparently, while data is distributed consistently in the background, adhering to privacy requirements.

\paragraph{Communication}

Finally, our edge compute infrastructure also provides abstractions for edge-to-edge communication, even between heterogeneous devices such as \glspl{uav} from different vendors.
Using the \gls{pubsub} paradigm, we provide a messaging middleware that integrates directly with the compute and data management layers of the edge infrastructure.
Unlike a cloud message bus, where all communication passes through the cloud, incurring high network overhead, our edge-to-edge message bus is decentralized and geo-distributed.
Prototypical implementations of such decentralized \gls{pubsub} middlewares exist with \emph{GeoBroker}~\cite{paper_hasenburg_geobroker,paper_hasenburg_disgb}, where messages are provided with geographical context and reach only local message subscribers.

\subsection{\acrshort{ai}-Assisted Communication System Approach}

In order to integrate split computing infrastructure with the communication network, the latter should be able to flexibly react to the computation demands of the applications on the one side, and to the environmental influences on the quality of provided communication links on the other side. The research focus in this field is the predictive algorithms for radio resource management, which would provide adjustable and guaranteed \gls{qos} parameters for communication participants.

\subsubsection{\acrshort{ai}-based system optimisation}

To meet the growing demands for performance and reliability in wireless communication systems, it is necessary to address the inherent uncertainty and dynamic nature of the wireless environment and radio access network. While these requirements cannot always be guaranteed,  \gls{ml} models can be used to adapt to the spatiotemporal dynamics of wireless networks in advance. By providing predictive analytics to both end-user applications and radio access network elements. \gls{ml} models can help ensure that these systems operate as efficiently and effectively as possible. Those are making it possible to make highly accurate predictions about changes in \gls{qos} and radio \glspl{kpi}, like \glspl{rem}, channel distribution information maps, and spectral efficiency, within the radio access network \cite{QZ_10012940,QZ_10012951,QZ_8813020,QZ_9354041,QZ_9569310}. These predictions can be used to support the proactive adaptation of end-user applications and optimize the radio access network in advance. 

By providing predicted \gls{qos} information, radio resource management schemes can offer more reliable future \gls{qos} guarantees to individual users, even when poor performance is expected. This means that the overall efficiency and reliability of wireless communication systems can be significantly improved with the help of \gls{ml}-powered predictive analytics. The characteristics of channel prediction methods on real communication data are also tested and verified by establishing an \gls{sdr} based cellular communication and channel measurement \cite{QZ_9044427,QZ_9860403}. The prediction uncertainty is unavoidable as a consequence of the stochastic nature and non-stationary of signal propagation associated with mobility. In addition, factors such as inefficient training data, increased feature space with uncertain inputs, and lack of generalizability in the trained \gls{ml} models further increase uncertainty in the predictions. 


Based on our implemented communication platform, \gls{ml} methods would be highly relied on to improve the throughput of the whole communication system and at the same time to prioritize well the users with poor expected performance in advance. 

\subsubsection{Agile Coding Schemes}

Offloading the computation tasks to the network in the approach of split computing creates complex applications, which are using the wireless interface for client-to-server communication. This leads to highly heterogeneous \gls{qos} requirements on the communication link since any application uses its own communication link. For certain applications, it might be advantageous to use several different links of data to tailor the \gls{qos} requirements to the particular data type. In order to provide the optimal experience for any type of data traffic, the wireless network should provide a high level of agility.

Fountain coding schemes could introduce further flexibility to a communication link. Fountain codes are rateless codes implementing $K$-out-of-$N$ decoding principle. Here, $N$ is the number of sub-codewords, which are generated from a data packet at the transmitter. $K$ is the least number of sub-codewords required to decode the packet at the receiver. Since there is no limitation for the number $N$ of generated sub-codewords, the coding rate can be flexibly adjusted. 
Fountain codes require an erasure channel, since they can withstand packet losses, but cannot recognise corrupted packets. Nevertheless, authors in  \cite{sm_berger2008optimizing, sm_karrenbauer2020study} showed, that combination of error-detecting codes with fountain codes gains the performance of the communication system.

Especially for real-time applications, the reliability of the communication link is highly important. Utilising \gls{ai} methods to predict the channel condition, the coding rate can be flexibly adopted to the varying radio environment in order to guarantee the required packet error rate \cite{sm_melnyk2022wireless}. This concept can be further extended to serve heterogeneous traffic classes, since the coding rate can be easily re-adjusted based on particular traffic requirements. Moreover, by means of fountain codes, distribution of the data over several communication links could be implemented in an agile way. Based on the state of every used link, the sub-codewords can be optimally distributed. If utilising different \glspl{rat}, the resiliency of the overall system would be improved. Furthermore, to improve the coverage area of a single \gls{rat}, a fully seamless handover could be realized by adaptively distributing the traffic between several communication links.

\subsubsection{Testbed System}
The project goals require approaches for processing functionality at several points in the infrastructure stack, which cannot be implemented in a public mobile communication network. The distributed processing intelligence has to be attached to the layers of cloud, edge and user equipment. 
For this purpose, transportable test stands for laboratory and open-field operation are being set up, which are available for use cases in laboratory or application-oriented environments.

In order to have comprehensive parameterisation and intervention options, the test stand setups are initially equipped with the OpenAirInterface software stack. As radio hardware, mixed \gls{sdr} components for base stations (gNBs) and terminals (UEs) from the USRP series by National Instruments and commercial modems in the \SIrange{3.7}{3.8}{\GHz} frequency band are used. This design should enable realistic latency and bandwidth testing of the use cases with a reduced number of subscribers compared to commercial installations.


For participation in flight testing in the \emph{Smart Drones} use case, a flight-capable terminal device is required that meets the application-related restrictions in terms of weight, size and power supply. In addition to multiconnective communication -- including participation in the \SIrange{3.7}{3.8}{\GHz} mobile radio network -- there should be options for connecting additional sensors for recording positions and movement states and wired communication to the flight controller.
This functionality is provided by a communication module derived from the \emph{NetMobilBox}~\cite{logicway_2021_NetMobilBox}, the development of which goes back to the \emph{5G NetMobil} project~\cite{logicway_2020_5GNetMobil}.

\section{Conclusion}

Sixth generation networks should bring mobile communication to the next level making the connectivity omnipresent for commercial and private usage. In order to realize this vision, new technological ideas are required. In this paper, we presented an approach and an architecture of joint communication and computing infrastructure. The proposed high performance backbone layer in combination with additional cloud native services allows the distribution of processing power in all levels inside the communication network, which reduces the demands on the client devices on the one hand, and simplifies the realization of collaborative applications on the other hand.

In order to support the development of the communication and computing infrastructure, two demanding use cases, \emph{Smart Drones} and \gls{hc}, were chosen. Those provide very heterogeneous requirements on latency, throughput, reliability, availability of computation power etc. Thus, this use cases provide appropriate 
testbed scenarios for the proposed infrastructure. The goal of the project \emph{6G NeXt} is to go beyond the conceptual evaluation of proposed technologies towards a functional proof of concept based on the proposed use cases.

\section*{Acknowledgements}


Funded by the Bundesministerium für Bildung und Forschung (BMBF, German Federal Ministry of Education and Research) -- 16KISK174K, 16KISK176, 16KISK177, 16KISK178, 16KISK179, 16KISK180, 16KISK182, 16KISK183.

\printbibliography

\balance



\end{document}